# Post-Fabrication Trimming of Silicon Photonic Ring Resonators at Wafer-Scale

Hasitha Jayatilleka, Harel Frish, Ranjeet Kumar, *Member, IEEE*, John Heck, *Member, IEEE*, Chaoxuan Ma, *Member, IEEE*, Meer Sakib, *Member, IEEE*, Duanni Huang, *Member, IEEE*, and Haisheng Rong, *Senior Member, IEEE, Fellow OSA*.

*Abstract*—Silicon ring resonator-based devices, such as modulators, detectors, filters, and switches, play important roles in integrated photonic circuits for optical communication, high-performance computing, and sensing applications. However, the high sensitivity to fabrication variations has limited their volume manufacturability and commercial adoption. Here, we report a low-cost post-fabrication trimming method to tune the resonance wavelength of a silicon ring resonator and correct for fabrication variations at wafer-scale. We use a Ge implant to create an index trimmable section in the ring resonator and an on-chip heater to apply a precise and localized thermal annealing to tune and set its resonance to a desired wavelength. We demonstrate resonance wavelength trimming of ring resonators fabricated across a 300 mm silicon-on-insulator (SOI) wafer to within +/-32 pm of a target wavelength of 1310 nm, providing a viable path to high-volume manufacturing and opening up new practical applications for these devices.

## I. INTRODUCTION

SILICON ring resonator-based devices are highly desirable in integrated photonics systems [1-4], especially microring modulators, for their small footprint, high-speed operation, and low-power consumption [3, 4]. Many of these benefits exhibited by ring resonators are due to their resonance characteristics which, unfortunately, also make them extremely susceptible to fabrication variations. Even typical nm-scale fabrication variations in the ring's geometry can significantly modify the roundtrip phase of the light inside the ring, causing its resonance wavelength to vary by a few nanometers (i.e., hundreds of gigahertz near 1310 nm wavelength) [5–9]. Therefore, the exact resonance wavelengths of ring resonator-based devices cannot be accurately designed and fabricated [10] and, hence, these devices require post-fabrication adjustments of their resonance wavelengths. In practice, thermo-optic phase tuners (i.e. heaters) are typically implemented to correct phase errors caused by fabrication variations at the cost of consuming a significant portion of the system's power budget [3, 5].

Alternatively, post-fabrication, non-volatile phase trimming allows corrections for fabrication variations by altering the refractive index of the waveguide or cladding materials after the devices are fabricated. While real-time tuning may still be required to account for any resonance drifts due to ambient temperature changes, post-fabrication trimming largely eliminates the power needed to correct for initial fabrication variations. Previous methods to implement changes to the refractive index include exposing fabricated devices to high-power laser beams [11–19], electron beams [6, 20, 21], or visible/UV light [14, 22]. These methods are difficult to implement at wafer-scale with high throughput and at low cost. More recently, using an on-chip heating element for post-fabrication trimming has shown promising results [23–26]. However, these demonstrations were performed at chip-level and have been limited by tuning range, trimming accuracy and controllability, or CMOS compatibility.

In this paper, we present a highly accurate, low-cost, and scalable method for post-fabrication resonance correction of silicon ring resonators at wafer-scale. This is achieved by making a small portion of the ring resonators' waveguides phase-trimmable by implanting with Ge and placing tungsten heaters above them for post-fabrication thermal annealing. The Ge implantation introduces disorders to the silicon lattice and changes the microstructure of the exposed regions of the silicon waveguides from crystalline to amorphous, altering the refractive index of the implanted regions [11]. Annealing these implanted regions at sufficiently high temperatures can cause the amorphized silicon in the waveguides to re-crystallize and change its refractive index back towards the original value. Therefore, by controlling the annealing temperature and duration using the integrated on-chip heaters, we can precisely adjust the refractive indices, and consequently, the rings' resonance wavelengths post fabrication. As we shall describe in this paper, this trimming process can be automated and applied to all ring resonators across entire SOI wafers to accurately tune their resonance wavelengths to desired values.

Hasitha Jayatilleka was with Intel Corporation, 2200 Mission College Blvd., Santa Clara, CA, 95054, USA (email: hasitha.jayatilleka@intel.com). Harel Frish is with Intel Corporation, 1600 Rio Rancho Blvd., Rio Rancho, NM, 871242, USA (email: harel.frish@intel.com). Ranjeet Kumar, John Heck, Chaoxuan Ma, Meer Sakib, Duanni Huang, and Haisheng Rong are with Intel Corporation, 2200 Mission College Blvd., Santa Clara, CA, 95054, USA (email: ranjeet.kumar@intel.com, john.heck@intel.com, chaoxuan.ma@intel.com, meer.nazmus.sakib@intel.com, duanni.huang@intel.com, haisheng.rong@intel.com).

## II. Device Design and Fabrication

The formation of the phase-trimmable silicon waveguide used in this work is illustrated in Fig. 1. The optical phase-change section is created by implanting a section of the silicon-on-insulator (SOI) rib waveguide with Ge through a photoresist mask [Fig. 1(a)]. Typical fabrication steps are then followed to deposit cladding oxide and to form a tungsten heater over the Ge implanted region of the waveguide section. The waveguide cross-section after fabrication is illustrated in Fig. 1(b). The transformation in the waveguide's microstructure can be visualized by the transmission electron microscope (TEM) images of implanted waveguides taken before and after annealing, shown in Figs. 1(c) and 1(d), respectively.

We estimate that the heaters used in this work can heat up the waveguide to 700 °C, well above the required temperatures of 450-600 °C for annealing out the implanted regions.

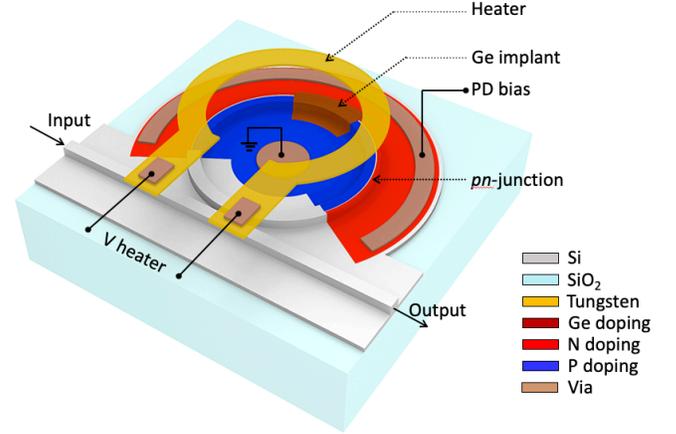

Fig. 2. Silicon microring modulator formed by a rib waveguide with an embedded pn-junction, a Ge doped section, and a metal heater above the ring. Top cladding oxide and metal contact pads are not shown.

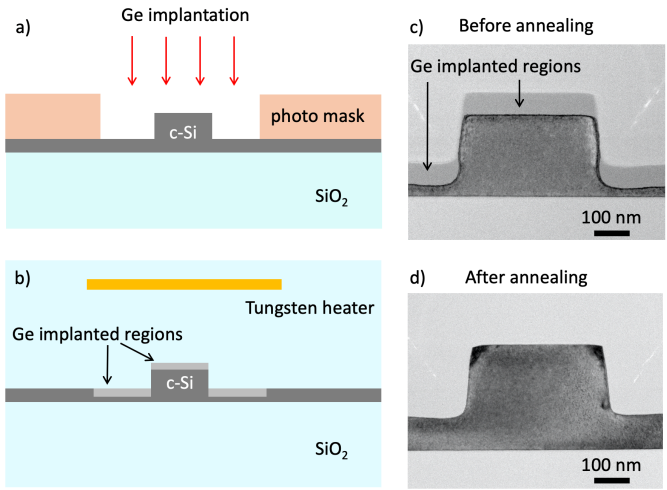

Fig. 1. Ge implanted silicon optical waveguide. a) Ge implantation on to a silicon waveguide over a photomask. b) Waveguide cross-section geometry after fabrication showing the Ge implanted phase-trimmable regions and the location of tungsten heater above these regions. TEM cross-sectional images of a fabricated waveguide c) before and d) after thermal annealing.

We used a microring optical modulator for a case study to demonstrate the viability of this phase trimming method on devices fabricated on a 300 mm SOI wafer in a standard CMOS fabrication facility. A schematic of the microring modulator with an integrated phase trimming section is shown in Fig. 2. The microring has a radius of 10 μm and is formed with a 400 nm wide rib waveguide with rib and slab heights of 300 nm and 100 nm, respectively. The Ge implant section is a 15 μm long, 2 μm wide arc overlapping the ring's waveguide. The tungsten heater of 40 μm long and 3 μm wide is curved along the ring.

We first studied the relationship between the roundtrip phase change, or the resonance wavelength shift, of the fabricated ring resonator under various thermal annealing conditions. We measured the resonance wavelengths of the ring resonator with a tunable laser and plotted the wavelength shifts as a function of the electrical heating power applied to the heater (Fig. 3). We increased the heating power in steps up to 550 mW and measured the resonance wavelength shift of the ring after each step. The ring was heated for a period of 5 mins at each step, then the heater was switched off, and the resonance wavelength was recorded. The temperature values shown in Fig. 3 were estimated by measuring the resonance shift of the ring resonator when the heater was on. We used a temperature coefficient of 63 pm/°C obtained in a separate experiment by measuring a ring's resonance shift as a function of the chip temperature. The free spectral range (FSR) of the ring near 1310 nm was measured to be 6.8 nm.

As shown in Fig. 3, no significant resonance wavelength shifts occurred when the heating power was below 350 mW. Beyond this point, however, the ring's resonance started to show a significant blue shift, indicating an optical phase change due to the thermal annealing. The total resonance blue shift of 8.5 nm corresponds to a -2.5π optical phase change or a change of the effective index $\Delta n_{eff} = -0.026$ of the waveguide.

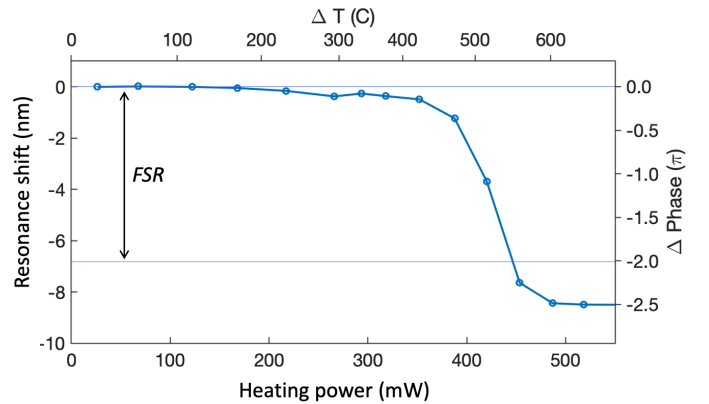

Fig. 3. Measured resonance wavelength shift of a microring modulator as a function of the applied electrical heating power. The corresponding phase change and the estimated temperature increase in the ring waveguide are also shown.

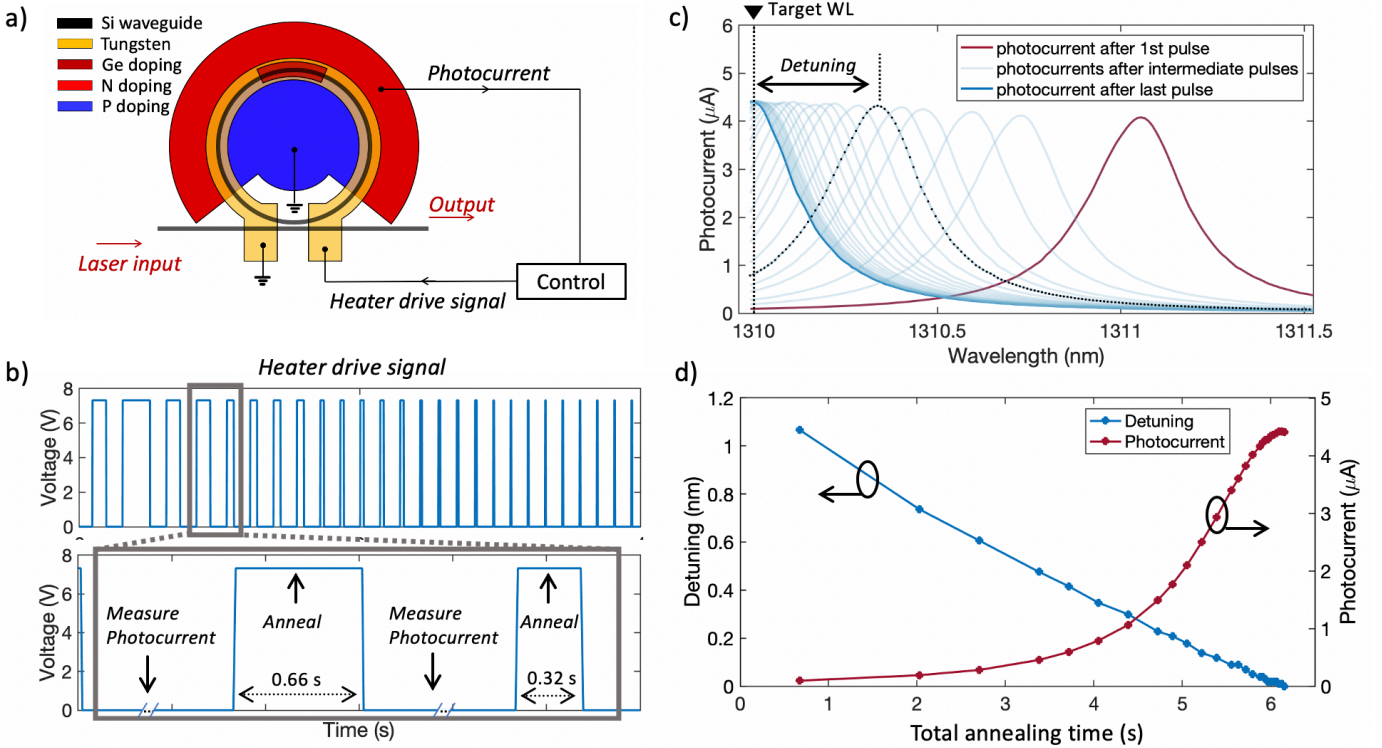

Fig. 4. Trimming the resonance of a microring modulator to a target laser wavelength. a) Illustration of the experimental configuration. b) Applied pulse width modulated (PWM) heater drive signal. Pulse width is dynamically adjusted in the feedback loop to gradually approach the target wavelength. c) Photocurrents measured after each voltage pulse by sweeping the laser during the off-periods of the PWM signal. The detuning from the target wavelength was calculated using these measured photocurrent curves (as shown). d) Measured resonance wavelength detuning from the target as a function of the total annealing time.

## III. Trimming Ring Resonator to a Target Wavelength

A $2.5\pi$ optical phase change is sufficient to trim the ring's resonance to any wavelength within its FSR of 6.8 nm. By controlling the heater's drive signal, we show that the ring's resonance wavelength can be trimmed to a desired wavelength target with high precision. Fig. 4(a) shows the experimental setup with the feedback loop for monitoring and trimming the resonance wavelength of the ring. We drive the heater with a pulse width modulated (PWM) voltage signal [Fig. 4(b)]. The PWM pulse height was 7.1 V, resulting in heating power of about 525 mW and a temperature increase of the waveguide underneath to about 650 °C for the duration of the pulse. Therefore, the ring was heated-up and annealed during the on-period of the pulse. When the ring cooled down to room temperature during the off-period, we measured the ring's resonance by sweeping the laser's wavelength and recording the photocurrent from the pn-junction. In this work, rather than using a separate photodetector (PD), we utilize the modulator's pn-junction as a PD for sensing the resonances of the microring modulator [27, 28].

Fig. 4(c) shows the recorded photocurrent curves from which we determined the ring's resonance wavelength detuning from the target of 1310 nm and used it as a feedback error signal to continuously adjust the duration of the subsequent pulses until the target wavelength is reached. The maximum pulse width was 1.2 s and the minimum pulse width was 0.01 s. After completing the trimming procedure, we measured the optical spectra of the device to confirm that the resonance has reached the target wavelength. Fig. 4(d) shows the measured detuning after each pulse as a function of the annealing time. The trimming speed was measured to be 0.18 nm/s as seen from Fig. 4(d).

In this proof-of-concept experiment, the total time used to trim the microring modulator was about 4 mins. This includes the time for laser sweeps, which we only performed to gain insights into the trimming process. However, as shown in Fig. 4(d), the total annealing time, i.e., the total on-time of the PWM drive signal, was less than 6.2 seconds. The time-consuming laser sweep can be avoided by first setting the input laser to the target wavelength and then trimming the ring until the photocurrent from the PD is maximized. As the thermal response time of the heater is typically in the order of 10s of microseconds, the off-period of the PWM signal can be reduced to less than a millisecond with a sufficiently fast PD read-out system. Therefore, the total trimming procedure would be dominated by the total annealing time of a few seconds.

After trimming, we applied a 275 mW constant heating power to the heater for 15 hours, which increased the temperature of the ring to about 300 °C. We did not observe any appreciable resonance movement under these conditions, reassuring that the resonance correction is permanent and would withstand typical packaging and operational temperatures. Furthermore, we note that the after-trimming stability of the devices can be improved by first annealing the wafer at a temperature of 325 °C for several minutes, or by applying serval burn-in pulses before the automated trimming procedure begins. This burn-in step would anneal out random defects and prevent permanent index changes that may occur at lower temperatures.

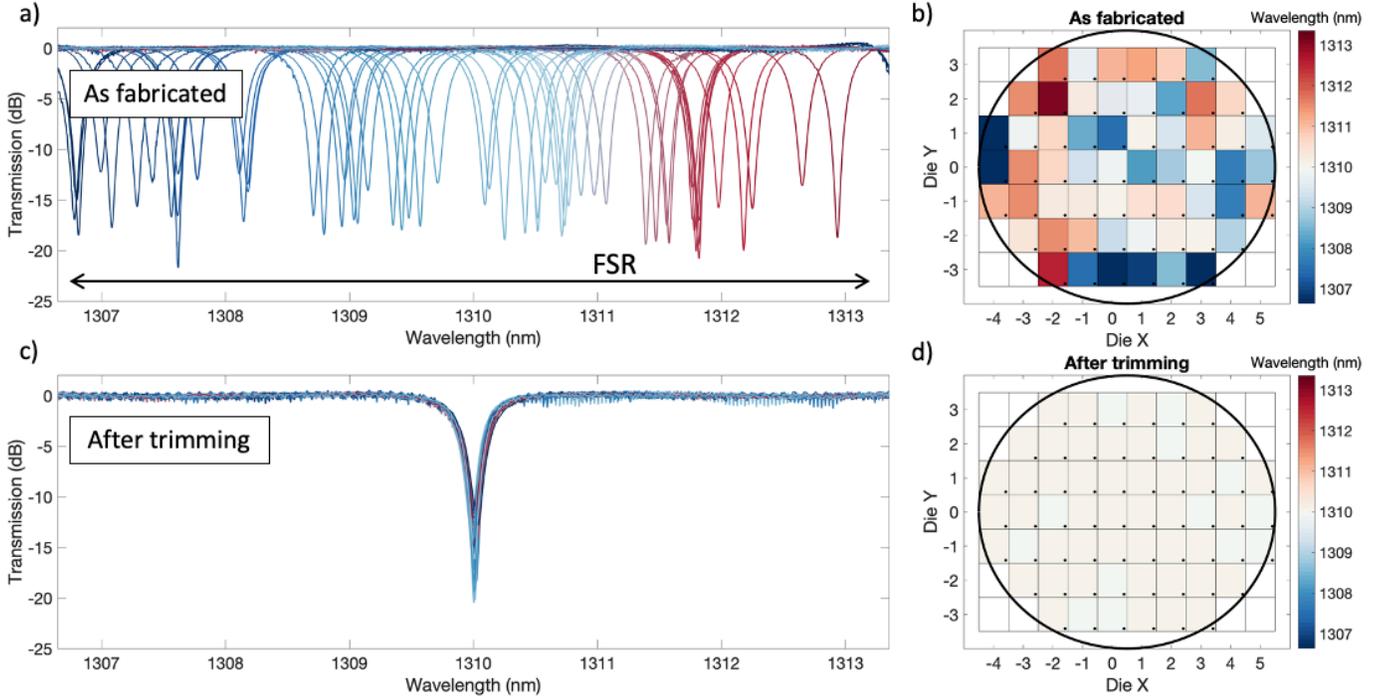

Fig. 5. Wafer-scale resonance trimming to 1310 nm target wavelength. As-fabricated a) transmission spectra spreading over an FSR and b) resonance wavelength distribution of a microring modulator at 58 die locations across a 300 mm wafer. c) Transmission spectra and d) resonance wavelength distribution of the microring modulators after trimming.

## IV. Wafer-Scale Trimming

Using this method, we performed automatic resonance wavelength trimming of microring modulators at wafer-scale. We fabricated the microring modulators, similar in design to that shown in Fig. 2, across 58 reticles on a 300 mm wafer. The as-fabricated transmission spectra of these devices and their locations across the wafer are shown in Figs. 5(a) and 5(b), respectively. As shown in Fig. 5(a), the as-fabricated resonance wavelengths of these microring modulators spread over more than a full FSR. We set the target wavelength to be 1310 nm, and trimmed all the ring resonators to this target using an automated wafer prober. After applying the described trimming procedure, the resonance wavelengths of all the rings were measured again and shown in Fig. 5(c) and 5(d). While the slow laser sweep increased the trimming time to several minutes per ring, the average annealing time required was 26.2 seconds per ring. This suggests that the total trimming time can approach this value with faster PD readout in a factory setting. After trimming, all of the resonances were measured to be within +/- 32 pm (5.6 GHz) of the target 1310 nm, with a standard deviation of $\sigma = 9$ pm. The mean quality (Q-) factor of the microring modulators after trimming was 4200, which corresponds to a photon lifetime limited modulation bandwidth of 55 GHz. The mean Q-factor of microring modulators similar in design but without the Ge implant was 6600. This reduction of Q corresponds to only 0.3 dB additional loss in the waveguide due to the 15 μm long Ge implant (see Appendix). The Ge doping concentration and annealing conditions can be further optimized in a future design to balance the trade-off between the residue loss and the achievable optical phase change. We also measured the modulation phase efficiency, $V_\pi L_\pi$, of the modulators' pn-junctions. After trimming, the mean $V_\pi L_\pi$ was 0.94 V·cm at a reverse bias of -1.5 V, as compared to a $V_\pi L_\pi$ of 0.79 V·cm of similar devices without the Ge implant. This change is likely due to the formation of amorphous silicon in the implant section of the pn-junction. This degradation in modulator's phase efficiency can be avoided by implanting Ge in the slab region of the waveguide, away from the pn-junction.

## V. Summary and Conclusion

In summary, we have demonstrated post-fabrication resonance wavelength trimming of silicon microring modulators at wafer-scale, opening up a viable path towards volume production and commercial adoption of silicon ring resonator-based devices. We have trimmed the resonance wavelengths of microring modulators across a 300 mm SOI wafer to a common wavelength target within +/- 32 pm, corresponding to an optical phase accuracy of $< 10^{-2}$ π. The trimming procedure was automated and was performed on the entire wafer using a wafer prober with an effective annealing time of 26.2 seconds per device. The trimming procedure could also be performed on packaged devices and/or on multiple devices simultaneously to maximize the throughput. The trimming accuracy and speed can be further improved by optimizing the pulse width and amplitude of the PWM drive signal while monitoring and tracking the pn-junction photocurrent from the microring modulators. After completing the post-process trimming, the tungsten heating element we use may also be driven under low voltage for thermal tuning to counteract real-time temperature fluctuations. While we have used microring modulators as a case study to demonstrate the viability of post-fabrication phase trimming in this paper, this trimming method is widely applicable. In addition to ring resonator-based devices [1-4, 28-30], it can generally be suitable for correcting fabrication errors in a wide range of phase-sensitive silicon photonics devices such as directional couplers [15], Mach-Zehnder interferometers [31-33], gratings [34, 35], and photonic crystal devices [17, 36]. This work represents a significant milestone in silicon photonics manufacturing by correcting optical phase errors post fabrication, removing a major hurdle for widespread adoption of ring resonator-based devices in commercial silicon photonics products.

APPENDIX

All microring modulators used grating couplers as optical inputs/outputs. The wavelength spectra shown in this paper, in Figs. 5(a), and 5(c), are normalized to the grating coupler's transmission spectra.

In order to obtain the excess loss from the Ge implant, we compared the implanted devices to non-implanted devices of otherwise identical design. The loss for each device is calculated by measuring the transmission spectra of the ring resonators and following methods described in [37]. After measuring all of the non-implanted and implanted devices across the wafer, the average excess loss from the 15 μm Ge implant is found to be 0.3 dB.

ACKNOWLEDGMENT

The authors would like to thank Saeed Fathololoumi, Ling Liao, Yuliya Akulova, and James Jaussi for technical discussions; Dina Unadkat, Brett Klehn, James Yonemura, and David Patel for device layout and testing support; and our colleagues in Intel Fab11x, Rio Rancho, NM, for device fabrication.